\documentstyle[12pt,aasms4]{article}
\begin{document}
\def\lapprox{\hbox{\lower .8ex\hbox{$\,\buildrel < \over\sim\,$}}}
\def\gapprox{\hbox{\lower .8ex\hbox{$\,\buildrel > \over\sim\,$}}}
\def\cago{$^{12}$C$(\alpha,\gamma)^{16}$O }
\def\O16{$^{16}$O}
\def\C12{$^{12}$C}
\def\XC{$X_{\rm C}$}
\def\XO{$X_{\rm O}$}

\title{The cooling of CO white dwarfs: influence of the internal 
chemical distribution}

\author{Maurizio Salaris\altaffilmark{1,2}, 
        Inmaculada Dom\'\i nguez\altaffilmark{3}, 
        Enrique Garc\'\i a-Berro\altaffilmark{4}, 
        Margarida Hernanz\altaffilmark{1}, 
        Jordi Isern\altaffilmark{1},
        Robert Mochkovitch\altaffilmark{5}}

\altaffiltext{1} {Institut d'Estudis Espacials de Catalunya,
                  CSIC Research Unit,
                  Edifici Nexus-104,
	          C/ Gran Capit\`a 2-4, 08034 Barcelona, Spain}
\altaffiltext{2} {Max-Planck-Institut f\"ur Astrophysik, Karl 
		  Schwarzschild  Strasse 1, 85740 Garching, Germany}
\altaffiltext{3} {Departamento de F\'\i sica Te\'orica y del Cosmos,
	          Universidad de Granada, Facultad de Ciencias,
	          18071 Granada, Spain}
\altaffiltext{4} {Departament de F\'\i sica Aplicada, Universitat
	          Polit\`ecnica de Catalunya, Jordi Girona Salgado s/n,
	          M\`odul B-5, Campus Nord, 08034 Barcelona, Spain}
\altaffiltext{5} {Institut d'Astrophysique de Paris, C.N.R.S., 98 bis 
		  Bd. Arago, 75014 Paris, France}

\received{} 
\accepted{}

\begin{abstract}

White dwarfs are the remnants of stars of low and intermediate masses
on the main sequence. Since they have exhausted all their nuclear fuel,
their evolution is just a gravothermal process. The release of energy 
only depends on the detailed internal structure and chemical composition 
and on the properties of the envelope equation of state and opacity;
its consequences on the cooling curve (i.e. the luminosity 
versus time relationship) depend on the luminosity at which this 
energy is released.

The internal chemical profile depends on the rate of the \cago reaction 
as well as on the treatment of convection. High reaction rates produce 
white dwarfs with oxygen rich cores surrounded by carbon rich mantles. 
This reduces the available gravothermal energy and decreases the 
lifetime of white dwarfs.

In this paper we compute detailed evolutionary models providing chemical 
profiles for white dwarfs 
having progenitors in the mass range from 1.0 to $7\,M_{\sun}$ and we 
examine the influence of such profiles in the cooling process. 
The influence of the process of separation of carbon and oxygen during 
crystallization is decreased as a consequence of the initial 
stratification, but it is still important and cannot be neglected. As 
an example, the best fit to the 
luminosity functions of Liebert et al. (1988) and Oswalt et al. (1996) 
gives and age of the disk of 9.3 and 11.0 Gyr, respectively, when this 
effect is taken into account, and only 8.3 and 10.0 Gyrs when it is 
neglected.

\end{abstract}


\section{Introduction}

The final result of the evolution of low and intermediate mass stars 
$(M \lapprox 7$--$8 M_{\sun})$ is a carbon-oxygen white dwarf. Since 
these stars have exhausted all their sources of nuclear energy, their 
evolution is determined by the gravothermal adjustement of their 
interiors induced by energy losses. As shown by Koester and 
Chanmugam (1990) using the virial theorem, this evolution can be 
interpreted in terms of a cooling process. The rate of cooling is 
determined, among other factors, by the ionic specific heat
which depends on the relative proportions of carbon and oxygen.
The change of chemical composition between the solid and the liquid at 
the onset
of crystallization and the gravitationally induced redistribution of 
carbon and oxygen provides an additional source of energy (Mochkovitch 
1983, Garc\'{\i}a-Berro et al. 1988), the importance of which depends, 
among other things, on the shape of the phase diagram.

Segretain et al. (1994) computed detailed cooling sequences using 
the most up to date input physics (both for the equation of state and 
the phase diagram), and taking into account both the release of latent 
heat and the release of gravitational energy induced by the 
redistribution of carbon and oxygen upon crystallization. The main 
result of including 
this extra energy source was a noticeable increase in the cooling 
times. For instance, the time taken by a typical 0.6 $M_{\sun}$ 
white dwarf with equal mass fractions of carbon and oxygen to 
reach a luminosity $\log(L/L_{\sun})=-4.5$ was 11.5 Gyr, when 
the redistribution process was properly taken into account, instead 
of 9.2 Gyr when this effect was neglected (a correction of $\sim$20\%). 

Two aspects that can reduce the efficiency of chemical redistribution 
arise from the assumptions that the liquid mantle surrounding the solid 
core is always perfectly mixed and that the white dwarf is initially 
made of half carbon and half oxygen uniformly distributed throughout 
the star. Although the validity of the first point was early studied 
by Mochkovitch (1983), it will be the object of an updated analysis in 
a forthcoming paper (Isern et al. 1996), and here we will concentrate 
on the role played by different chemical initial profiles.

Mazzitelli \& D'Antona (1986a,b; 1987) studied the evolution from
the main sequence to the white dwarf stage for 1, 3 and 5 $M_{\sun}$
stars and found that the composition (and in particular the
carbon-oxygen ratio) of the resulting white dwarf was very
sensitive to the adopted cross section for the \cago reaction,
to the detailed prescriptions adopted for convective mixing,
and to the mass of the star on the main sequence. Their results
indicate that a typical 0.6 $M_{\sun}$ white dwarf consists of 
an inner oxygen-rich core surrounded by a carbon-rich mantle, 
whereas a 1.0 $M_{\sun}$ white dwarf has almost flat carbon-oxygen 
profiles with roughly \XC=\XO=0.5 (by mass). This has two effects: 
first, since oxygen crystallizes at higher temperatures than carbon, 
both the latent heat and the gravitational energy are released at 
higher luminosities and the induced delay in the cooling times is smaller. 
Second, since the inner core is oxygen-rich the gravitational energy 
released upon crystallization is smaller, thus reducing the effect 
of the redistribution process. The profiles obtained by 
Mazzitelli \& D'Antona (1986a,b) were also adopted 
by Segretain et al. (1994), who found that the time taken by a 
0.6 $M_{\sun}$ white dwarf to reach 
$\log(L/L_{\sun})=-4.5$ was 8.8 Gyr or 10.0 Gyr, depending on 
whether 
the redistribution process was neglected or taken into account. 
Thus, the delay induced by phase separation was 1.2 Gyr, 
which is a correction of $\sim$15\%.

The purpose of this paper is to clarify, in light of the new
determinations of the \cago reaction rate, whether or not the
interiors of white dwarfs are stratified before crystallization
sets in and to determine the effect of the actual chemical profile 
on the cooling times, thereby providing better estimates of the 
ages of white dwarfs and of the age of the solar neighborhood.

\section{Input physics}

The evolutionary stellar models presented in this paper have
been computed using the evolutionary code FRANEC (Frascati RAphson
Newton Evolutionary Code), as described in Chieffi \& Straniero
(1989); the reader is referred to this paper for an exhaustive
discussion about the physical inputs adopted in the code. In the
following we will discuss briefly only some basic features relevant  
for the scope of this paper.

The boundaries of convective regions are set by adopting the
Schwarzschild criterion and no mechanical overshooting is allowed.
Semiconvection during central helium burning is computed according
to the method described in Castellani et al. (1985), and the breathing
pulses occuring during the last portion of core helium 
burning have been inhibited. For $T>10^4$ K, the OPAL radiative 
opacities of Iglesias et al. 
(1992) were used, whereas for $T \leq 10^4$ K, the opacities of 
Kurucz (1991) were adopted.
We have assumed a value $Z=0.02$ for the solar metallicity and the
heavy elements distribution as derived by Grevesse (1991). The solar
helium abundance $Y_{\sun}$ and the value of the mixing length
parameter $\alpha$ have been derived by matching the luminosity
and radius of a stellar model with $Z=0.02$ and the solar age to
their solar values. The values obtained are $\alpha=2.25$ and
$Y_{\sun}=0.289$ --- see Chieffi, Straniero \& Salaris (1995) for
a detailed discussion about the solar calibration with the new
OPAL and Kurucz opacities.

Nuclear reaction rates have been taken from Fowler, Caughlan \&
Zimmerman (1975), and the subsequent modifications have been taken
from Harris et al. (1983), Caughlan et al. (1985) and Caughlan \&
Fowler (1988). The reaction rate for the \cago reaction is crucial 
for this study, since during the He burning phase, when central 
helium is depleted down to $Y=0.10$, the burning mainly occurs 
through this reaction, and its rate determines the \C12 and \O16 
profiles in the final white dwarf structure. 

Several different teams have recently examined the cross section of the 
\cago reaction (Ji et al. 1990, Zhao et al. 1993, Buchmann et al. 
1993, Azuma et al. 1994, Mohr et al. 1995, Trautvetter 1996). A 
detailed analysis of the data obtained shows that these experiments
are compatible with a total astrophysical S-factor at 300 keV 
$(S_{300})$ in the range 120\, keV~b \lapprox $S_{300}$\lapprox 
220\,keV~b (Trautvetter 1996). 
There have also been several attempts to constrain the \cago reaction
rate from astrophysical data. However, the fractions of \C12 and \O16 
produced in a typical star depend both on the reaction rate and on the 
treatement of convection. Thus, since we cannot disentangle both 
effects, these constraints are only set on an {\sl effective} cross 
section for the \cago reaction, given the lack of a reliable theory of
convection.

Woosley, Timmes \& Weaver (1993) studied the role of the \cago 
reaction rate in producing the solar abundance set from stellar 
nucleosynthesis and concluded that the effective astrophysical 
S-factor for the energies involved during core helium burning 
that best reproduces the observed abundances should be $S_{300}=170$ 
keV~b, in good agreement with the experimental data. This value for 
$S_{300}$ corresponds to the value given by Caughlan \& Fowler (1988) 
multiplied by $\sim$1.3. In their 
models the Ledoux criterion plus an amount of convective overshooting 
were adopted for determining the extension of the convective regions. 

Thielemann, Nomoto \& Hashimoto (1996) studied the collapse of 
gravitational supernovae and compared the predicted amount of 
\C12 and \O16 in their ejecta with the abundances observed 
in SN1987A and SN1993J. They found that the agreement was 
excellent when the reaction rate of Caughlan et al. (1985), 
which was computed assuming $S_{300}=240$ keV~b, and the 
Schwarzschild criterion without overshooting were adopted.

Therefore, given our treatment of convection, we have adopted the rate 
of Caughlan et al. (1985) for the \cago reaction. However, for a sake 
of comparison, two evolutionary sequences producing the most probable 
final white dwarf configurations --- that is white dwarf masses between 
0.55 and 0.65 $M_{\sun}$ --- have also been computed using a lower 
cross section, namely the rate inferred by Woosley, Timmes \& Weaver (1993) 
($S_{300}=170$ keV~b, see above), hereinafter ``low rate".

\section{The properties of the CO cores}

With the input physics briefly described above, we have computed
evolutionary sequences --- neglecting mass loss --- from the zero
age main sequence to the thermally pulsing asymptotic giant branch
(AGB) phase, of stellar models with masses in the range $1.0\leq 
M/M_{\sun}\leq 7.0$. The tracks in the Hertzsprung-Russell diagram 
of these model sequences are shown in 
Figure 1, each one labeled with its corresponding mass. 

\placefigure{fg1}

The evolutionary sequences were terminated 
at the end of the first thermal pulse. A real counterpart of these 
models is expected to lose its hydrogen-rich envelope during the 
thermally pulsing AGB phase, thanks to a rapid radiative wind, before 
becoming the central star of a planetary nebula, and finally evolve into 
a white dwarf. We have not explicitly followed these evolutionary 
phases since we are only interested in the chemical composition 
of the CO core and in the initial-final mass relation, in order to 
compute white dwarf cooling sequences (see below for details). As 
these phases can affect the initial-final mass relation (but not 
the \C12 and \O16 chemical profiles within the core), we have checked that 
our values for the mass internal to the He-H discontinuity $(M_{\rm WD})$ 
as a function of the main sequence mass $(M_{\rm MS})$ --- see Table 1 ---
are compatible with the semi-empirical relation given by Weidemann 
(1987) and with that of Iben \& Laughlin (1989). 
The maximum difference between our results and these two relations is of 
the order of 10\%, well within the uncertainty associated to the 
semi-empirical determination.
The time spent during the pre-white dwarf phase $(t)$ and the oxygen 
abundance at the center (\XO) at the end of the first thermal pulse are
also displayed in Table 1.
The maximum difference between our evolutionary times and those published
by Iben \& Laughlin (1989) --- their equation 22 --- is 1\% in $\log t$.

\placetable{table1}

Figure 2 displays the oxygen profiles for some of the CO cores obtained 
just at the end of the first thermal pulse \footnote{Detailed chemical 
abundance profiles for the CO cores described in this section are 
available upon request to the authors.}. 
The inner part of the 
core, with a constant abundance of \O16, is determined by the maximum 
extension of the central He-burning convective region while the peak in 
the oxygen 
abundance is produced when the He-burning shell crosses the semiconvective 
region partially enriched in \C12 and \O16, and carbon is converted 
into oxygen through the \cago reaction. 
Beyond this region, the oxygen profile is built when the thick 
He-burning shell 
is moving towards the surface. Simultaneously, gravitational contraction 
increases its temperature and density, and since the ratio between the 
\cago and $3\alpha$ reaction rates is lower for larger temperatures 
--- see e.g. Figure 1 in Mazzitelli \& D'Antona (1987) --- the oxygen 
mass fraction steadily decreases in the external part of the CO core.
It is also interesting to notice that, 
in contrast with Mazzitelli and D'Antona (1987), all the models, 
including those of the highest mass, have their central regions dominated 
by oxygen (see Table 1). However, the amplitude of the peak in the 
oxygen abundance profile decreases as the mass of the CO core increases.
Figure 2 also displays the \O16 profile (dotted line) of the white 
dwarf resulting from the evolution of a 3.2 $M_{\sun}$ stellar model, 
computed adopting the low value of the \cago reaction rate. The shape 
of the 
chemical profile is similar to that obtained by adopting the rate of 
Caughlan et al. (1985), but now, due to the less efficient conversion 
of \C12 into \O16, the two elements have a more similar abundance in 
the inner part of the core (\XC=0.40, \XO=0.57).

\placefigure{fg2}

The \C12 and \O16 profiles at the end of the first thermal pulse 
have an off-centered peak in the oxygen profile,  which is related 
to semiconvection (as explained above). Since we have chosen the 
rate of Caughlan et al. (1985) for the \cago reaction, we were 
forced to use the Scwarzschild criterion for convection (see the 
previous discussion in section 2) and, therefore, we did not find 
the chemical profiles to be Rayleigh-Taylor unstable during the early 
thermally-pulsing AGB phase. After the ejection of the envelope, 
when the nuclear reactions are negligible at the edge of the degenerate 
core, the Ledoux criterion can be used and, therefore, the chemical 
profiles are Rayleigh-Taylor unstable and, consequently, will be 
rehomogeneized  by convection (Isern et al. 1996). Notice that, in any 
case, this rehomogeneization minimizes the effect of the separation 
occurring during the cooling process.
Figure 3 shows the oxygen profile obtained for the $3.2\,M_{\sun}$ 
model at the end of the first thermal pulse (dotted line), and 
the resulting profile after rehomogeneization (dotted-dashed line).
The resulting profiles after Rayleigh-Taylor rehomogeneization are 
the initial profiles adopted in our cooling sequences.

\placefigure{fg3}

\section{White dwarf cooling ages and luminosity functions}

We have computed cooling sequences for the carbon-oxygen cores
previously described according to the method developed by D\'\i az-Pinto 
et al. (1994) and subsequently modified in Garc\'\i a--Berro et al. (1996). 
This method assumes that the white dwarf has an isothermal core and that 
the luminosity is only a function of its mass and temperature. 
The adopted relationship between the core temperature and the 
luminosity is a fit to the results of Wood \& Winget (1989) for a 
0.6 $M_{\sun}$ CO white dwarf, with 
a helium envelope of mass $10^{-4}\,M_{\rm WD}$, conveniently scaled by mass, 
which is enough for our purposes.
The cooling times and the characteristic cooling timescales are derived 
from the
binding energy of the white dwarf and the aforementioned relationship.
Our cooling sequences start at core temperatures of $5\times 10^7$ K, 
which roughly correspond to luminosities $\sim 10^{-1}\,L_{\sun}$ 
--- well below the knee in the Hertzsprung-Russell diagram. 
Neutrino cooling at high temperatures (i.e. high luminosities) has 
been included as in Garc\'\i a-Berro et al. (1996). We have used 
the equation of state described in Segretain et al. (1994) which 
includes accurately all the relevant contributions to the 
thermodynamical quantities both in the liquid and in the solid 
phase. Phase separation during solidification has been included, using 
the phase diagram of the carbon-oxygen binary mixture of Segretain \& 
Chabrier (1993), which is of the spindle form. 

During the crystallization process of the white dwarf interior, 
the chemical composition of the solid and liquid phases are 
not equal. A solid, oxygen-rich core grows and the lighter 
carbon-rich fluid which is left ahead of the crystallization 
front is Rayleigh-Taylor unstable and is efficiently redistributed 
by convective motions in the outer liquid mantle (see Appendix 
A). The net effect is a migration of some oxygen towards the central 
regions which leads to a subsequent energy release (Mochkovitch 1983, 
Isern et al. 1996). The final profile for a 0.61 $M_{\sun}$ white dwarf,  
when the whole interior has crystallized, is shown in Figure 3 as a solid 
line.

The cooling times as a function of the luminosity 
for the different models computed are shown in Table 2
and in Figure 4 
\footnote{Detailed cooling sequences are available upon request  
to the authors.}. 
The onset of crystallization is clearly marked by the change in
the slope of the cooling curves. Obviously, massive white dwarfs 
crystallize at higher temperatures (luminosities) because they 
have larger central densities and their oxygen abundance does not 
vary significantly.
As an example of the influence of phase separation in the cooling
times, the time taken
by a 0.61 $M_{\sun}$ white dwarf to reach $\log (L/L_{\sun})=-4.5$ is
9.9 Gyr (to be compared with 8.9 if phase separation is neglected). 
For comparison we have also computed the cooling sequence for the 
0.6 $M_{\sun}$ white dwarf obtained 
using the low \cago rate (see Figure 2 for the \O16 profile). 
The time necessary to reach $\log(L/L_{\sun}) \simeq -4.5$, 
is now 10.3 Gyr (to be compared with 9.2 Gyr if separation is 
neglected).
Two aspects of these latter results deserve further discussion. 
First, either if phase separation is neglected or not, the cooling 
ages are larger for the low rate model sequence. This is due to 
their lower oxygen content, which leads to a larger heat capacity and, 
therefore, to a slower cooling rate. Second, the delay introduced by 
phase separation during crystallization down to 
$\log (L/L_{\sun})=-4.5$ is practically the same for both cooling 
sequences 
($\sim$ 1 Gyr). The reason for this is twofold: on one hand, in 
the model 
computed with the low rate of the \cago reaction, less oxygen is 
available for separation ($M_{\rm O} = 0.3 M_{\sun}$ instead of 
$M_{\rm O} = 0.4 M_{\sun}$, see Figure 2) but, on the other hand, the 
change in the oxygen abundance upon crystallization is larger, due 
to the spindle form of the phase diagram --- see Figure 2 in Segretain 
et al. (1994).

In order to compare our new results for the cooling times with those
of our previous works (Segretain et al. 1994, Hernanz et al. 1994),
we should take into consideration the two basic improvements  
introduced since then. First of all, the
profiles of chemical composition for the different possible progenitors
of carbon-oxygen white dwarfs have been obtained using the best available
effective rates for the \cago reaction and updated input physics. 
For a typical 0.6 $M_{\sun}$ white dwarf, 
the total oxygen mass is similar to that found by Mazzitelli \& 
D'Antona (1986b), $M_{\rm O} = 0.4\,M_{\sun}$, but it is distributed in 
a different way. And second,
we have improved the treatment of solidification,
by taking into account the actual profile instead of approximating it
by a two-step function (see Segretain et al. 1994 and Garc\'{\i}a-Berro
et al. 1996). Our present treatment is more realistic and further reduces 
the effect of phase separation. 
We have recomputed the cooling sequence of
a $0.6\,M_{\sun}$ white dwarf using the exact initial chemical profile 
of Mazzitelli \& D'Antona (1986b); the delay introduced by phase 
separation in the cooling time down to $\log(L/L_{\sun}) \simeq -4.5$ 
is now 0.9 Gyr instead of the previous 1.2 Gyr found in Segretain et al. 
(1994), which is similar to the value obtained with the chemical 
profiles derived in \S3 (1 Gyr).

\placetable{table2}
\placefigure{fg4}

Another important magnitude related to the cooling, which is directly
involved in the calculation of the white dwarf luminosity function, is 
the characteristic cooling timescale, defined as
$\tau_{\rm cool} = dt_{\rm cool}/dM_{\rm bol}$. This quantity
is shown in Figure 5 as a function of the luminosity, for the
masses listed in Table 2.
The onset of crystallization is clearly marked by a sudden increase of 
the $\tau_{\rm cool}$ versus $\log(L/L_{\sun})$ relation, which corresponds 
to the change in the slope of the cooling curves. During the 
solidification of their interiors, white dwarfs 
must radiate away both the extra amount of energy due to the
release of
the latent heat of crystallization and the gravitational 
energy released by phase separation, thus slowing down the cooling 
process and consequently increasing the characteristic cooling 
timescales. The amplitude of the bump is smaller for massive 
white dwarfs because the release of energy at crystallization 
occurs at higher luminosities. 

\placefigure{fg5}

Finally, in order to examine the influence of our cooling sequences 
on the estimation of the age of the solar neighborhood, we have computed 
white dwarf luminosity functions (see Figures 6 and 7), with 
the method explained in Hernanz et al. (1994), assuming a
Salpeter-like initial mass function (Salpeter 1961) and a constant star 
formation rate per unit volume.
The age of the disk that best fits the observational
data of Liebert et al. (1988), when adopting blackbody corrections for
the cool non-DA white dwarfs (see figure 6), is 9.3 Gyr --- see Hernanz 
et al. (1994) for a discussion of the uncertainty of the age determination 
associated to the exact position of the cutoff \footnote{For a general 
discussion, within the context of galactic evolution, of the uncertainties 
of age determinations using the white dwarf luminosity function, see Isern 
et al. (1995a, b) and Wood (1992).}. 
This age of the disk has to be compared with an age of
8.3 Gyr, obtained using the same set of inputs but neglecting 
phase separation.
If the observational data set of Oswalt et al. (1996) is adopted (see
Figure 7), our 
best fit corresponds to an age of 11.0 Gyr (10.0 Gyr if phase separation
is neglected)

\placefigure{fg6}
\placefigure{fg7}

\section{Conclusions}

In this paper we have examined the influence of the \cago reaction
rate on the final structure of carbon-oxygen white dwarfs.
The FRANEC evolutionary code, with updated input physics, has
been used to derive accurate chemical profiles. The full range of 
initial
masses producing carbon-oxygen white dwarfs as a final result, 
in the frame of single star evolution, has been analyzed. For the best 
choice of the combined effect of convection and the \cago reaction 
rate, carbon-oxygen profiles showing an enhancement of oxygen 
in the central regions for all core masses are obtained, whereas
for a lower \cago reaction rate, this effect is smaller. 
Mass fractions of \O16 as high as 0.83 are obtained for a 
0.55 $M_{\sun}$
white dwarf, descending from a 2.5 $M_{\sun}$ star, whereas 
$X_{\rm O}=0.66$ 
corresponds to a 1.0 $M_{\sun}$ white dwarf, descending from a 
7 $M_{\sun}$ star.

The resulting 
carbon-oxygen profiles have been used for computing white dwarf
cooling sequences, including the effect of phase separation during
solidification. This phenomenon leads to a non negligible increase of 
the cooling ages,
which translates into an increase of the age of the disk
of the same order. Our best estimate of
the age of the disk is 9.3 Gyr, when the data set of Liebert et al. 
(1988) is used, 
and 11.0 Gyr when the data set of Oswalt et al. (1996) is adopted, 
in contrast with 8.3 and 10.0 Gyr, obtained, respectively, when phase 
separation is neglected. These values 
indicate that the effects associated with crystallization should not 
be neglected when using white dwarfs as a tool to determine the age of 
the disk.

\acknowledgements
	This work has been supported by DGICYT grants PB94-0111 and
	PB94--0827-C02-02, by the CIRIT grant GRQ94-8001, by the 
	AIHF 335-B, by the AIHI 94-082-A and by the C$^4$ consortium. 
	One of us (M.S.) thanks the E.C. for the ``Human Capital 
	and Mobility'' fellowship ERBCHGECT920009.

\newpage

\appendix{Appendix A: Change of the chemical profile during the
solidification process}

The distribution of carbon and oxygen in the outer liquid mantle
of a crystallizing white dwarf is Rayleigh-Taylor unstable because a
lighter carbon-rich fluid is released at the crystallization boundary.
Convective mixing will redistribute the abundances and lead to flat
profiles in a region whose size depends on the initial composition 
profile and on the degree of enrichment produced during the 
solidification process.

Consider, therefore, a partially solidified white dwarf of total 
mass $M_{\rm WD}$ containing a total amount of oxygen $M_{\rm O}$. Its 
structure can be divided into three parts: a solid core of mass 
$M_{\rm S}$ and oxygen mass fraction $X_{\rm S}(m)$, a liquid mantle of 
mass $\Delta M$ homogenized by convection, with oxygen abundance $X$, 
and an outer, 
unperturbed region, with the initial oxygen profile $X_{\rm O}(m)$.
Therefore, the total mass of oxygen can be written as

\begin{equation}
M_{\rm O}=\int_0^{M_{\rm S}}X_{\rm S} dm + X(M_{\rm L}-M_{\rm S}) 
+ \int_{M_{\rm L}}^{M_{\rm WD}} X_{\rm O} dm \eqnum{A.1}
\end{equation}

\noindent 
where $M_{\rm L}=M_{\rm S}+\Delta M$

After deriving this expression with respect to the solid mass and 
introducing the two conditions $X_{\rm S}(M_{\rm S})=(1+\alpha)X$ 
and $X=X_{\rm O} (M_{\rm S})$, where $\alpha$, which depends on $X$, 
is the degree of enrichment produced during crystallization, we 
obtain:

\begin{equation}
\alpha X +\frac{dX}{dM_{\rm S}}(M_{\rm L}-M_{\rm S})=0 \eqnum{A.2}
\end{equation}

\noindent
introducing $q=M_{\rm S}/M_{\rm WD}$ and $q_{\rm L}=M_{\rm L}
/M_{\rm WD}$,

\begin{equation}
\frac{dX}{dq}[q_{\rm L}(X)-q]+\alpha(X) X=0\eqnum{A.3}
\end{equation}

\noindent
Notice the singularity at $q=0$ since $q_{\rm L}(X)=0$ and also notice 
that if the initial profile is flat, $q_{\rm L}(X)=1$.
Integrating this equation provides the final oxygen profile after 
crystallization.

\newpage

\figcaption[f1.ps]{Evolutionary tracks in the Hertzsprung-Russell 
diagram for the model sequences quoted in the text. Each track is 
labeled with its corresponding mass.
\label{fg1}}

\figcaption[f2.ps]{Oxygen profiles for selected white dwarf models 
with masses of 0.61, 0.68 and 0.87 $M_{\sun}$ and our choice of the 
\cago reaction rate (solid line). The dotted 
line displays the oxygen profile for the 0.60 $M_{\sun}$ model computed 
by adopting the low rate of the \cago reaction. 
\label{fg2}}

\figcaption[f3.ps]{Oxygen profile of a 0.61 $M_{\sun}$ white dwarf at 
the beginning of the thermally-pulsing AGB phase (dotted line), 
the same after rehomogenization by Rayleigh-Taylor instabilities 
during the liquid phase (dotted-dashed line) and after total 
freezing (solid line). 
\label{fg3}}

\figcaption[f4.ps]{Cooling curves (time is in Gyr) for the white dwarf models 
described in the text and in Table 1 (except for the ``low rate" case).
\label{fg4}}

\figcaption[f5.ps]{Characteristic cooling times (in yr) for the same models 
shown in Figure 4.
\label{fg5}}

\figcaption[f6.ps]{Luminosity function obtained assuming a constant 
star formation rate per unit volume and an age of the disk of 9.3 Gyr. 
The observational data are from Liebert, Dahn and Monet (1988).
\label{fg6}}

\figcaption[f7.ps]{Luminosity function obtained assuming a constant 
star formation rate per unit volume and an age of the disk of 11.0 Gyr. 
The observational data are from Oswalt, Smith, Wood and Hintzen (1996).
\label{fg7}}

\newpage

\begin{deluxetable}{cccc}
\tablecaption{Characteristics of the white dwarfs obtained adopting
the rate of Caughlan et al. (1985) for the \cago reaction 
\label{table1}}
\tablehead{
\colhead{$M_{\rm MS} (M_{\sun})$} &
\colhead{$\log t$ (yr)} &
\colhead{$M_{\rm WD} (M_{\sun})$} &
\colhead{\XO}
}
\tablewidth{0 pt}
\startdata
1.0 & 10.0 & 0.54 & 0.79\nl
1.5 & 9.36 & 0.54 & 0.79\nl
2.0 & 9.02 & 0.54 & 0.79\nl
2.5 & 8.88 & 0.55 & 0.83\nl
3.2 & 8.56 & 0.61 & 0.74\nl
3.6 & 8.41 & 0.68 & 0.72\nl
4.0 & 8.27 & 0.77 & 0.71\nl
5.0 & 8.01 & 0.87 & 0.68\nl
7.0 & 7.66 & 1.00 & 0.66\nl
\enddata
\nl
\end{deluxetable}

\newpage
\begin{deluxetable}{ccccccccc}
\footnotesize
\tablecaption{Cooling times, in Gyr, for white dwarfs of different masses.
\label{table2}}
\tablehead{
\colhead{$-\log(L/L_{\sun}$)} & & & &$t_{\rm cool}$ & & & \\ 
\cline{2-9} &
\colhead{ 0.54 $M_{\sun}$ } &
\colhead{ 0.55 $M_{\sun}$ } &
\colhead{ 0.61 $M_{\sun}$ } &
\colhead{ 0.60 $M_{\sun}$ \tablenotemark{a}} &
\colhead{ 0.68 $M_{\sun}$ } &
\colhead{ 0.77 $M_{\sun}$ } &
\colhead{ 0.87 $M_{\sun}$ } &
\colhead{ 1.00 $M_{\sun}$ } 
}
\tablewidth{0 pt}
\startdata
2.00 & 0.10 & 0.10 & 0.12 & 0.12 & 0.16 & 0.21 & 0.28 & 0.42 \nl
2.20 & 0.17 & 0.18 & 0.22 & 0.22 & 0.27 & 0.33 & 0.42 & 0.60 \nl
2.40 & 0.28 & 0.29 & 0.34 & 0.34 & 0.40 & 0.48 & 0.60 & 0.81 \nl
2.60 & 0.42 & 0.43 & 0.48 & 0.49 & 0.56 & 0.67 & 0.80 & 1.06 \nl
2.80 & 0.58 & 0.60 & 0.67 & 0.68 & 0.76 & 0.89 & 1.05 & 1.39 \nl
3.00 & 0.79 & 0.81 & 0.89 & 0.91 & 1.02 & 1.17 & 1.37 & 1.98 \nl
3.20 & 1.05 & 1.08 & 1.18 & 1.21 & 1.34 & 1.54 & 1.89 & 2.81 \nl
3.40 & 1.39 & 1.43 & 1.57 & 1.60 & 1.77 & 2.11 & 2.79 & 3.83 \nl
3.60 & 1.85 & 1.90 & 2.09 & 2.14 & 2.58 & 3.16 & 4.02 & 5.03 \nl
3.80 & 2.66 & 2.73 & 3.19 & 3.02 & 3.91 & 4.68 & 5.57 & 6.36 \nl
4.00 & 4.09 & 4.22 & 4.86 & 4.91 & 5.72 & 6.58 & 7.30 & 7.67 \nl
4.20 & 5.94 & 6.20 & 6.87 & 7.06 & 7.80 & 8.50 & 8.86 & 8.69 \nl
4.40 & 7.98 & 8.30 & 9.01 & 9.22 & 9.60 &10.00 & 9.99 & 9.40 \nl
4.50 & 8.88 & 9.22 & 9.85 &10.26 &10.33 &10.59 &10.45 & 9.69 \nl
4.60 & 9.69 &10.02 &10.60 &11.25 &10.97 &11.12 &10.86 & 9.95 \nl
4.70 &10.43 &10.75 &11.27 &12.21 &11.56 &11.61 &11.25 &10.20 \nl
\enddata
\tablenotetext{a}{low rate}
\nl
\end{deluxetable}

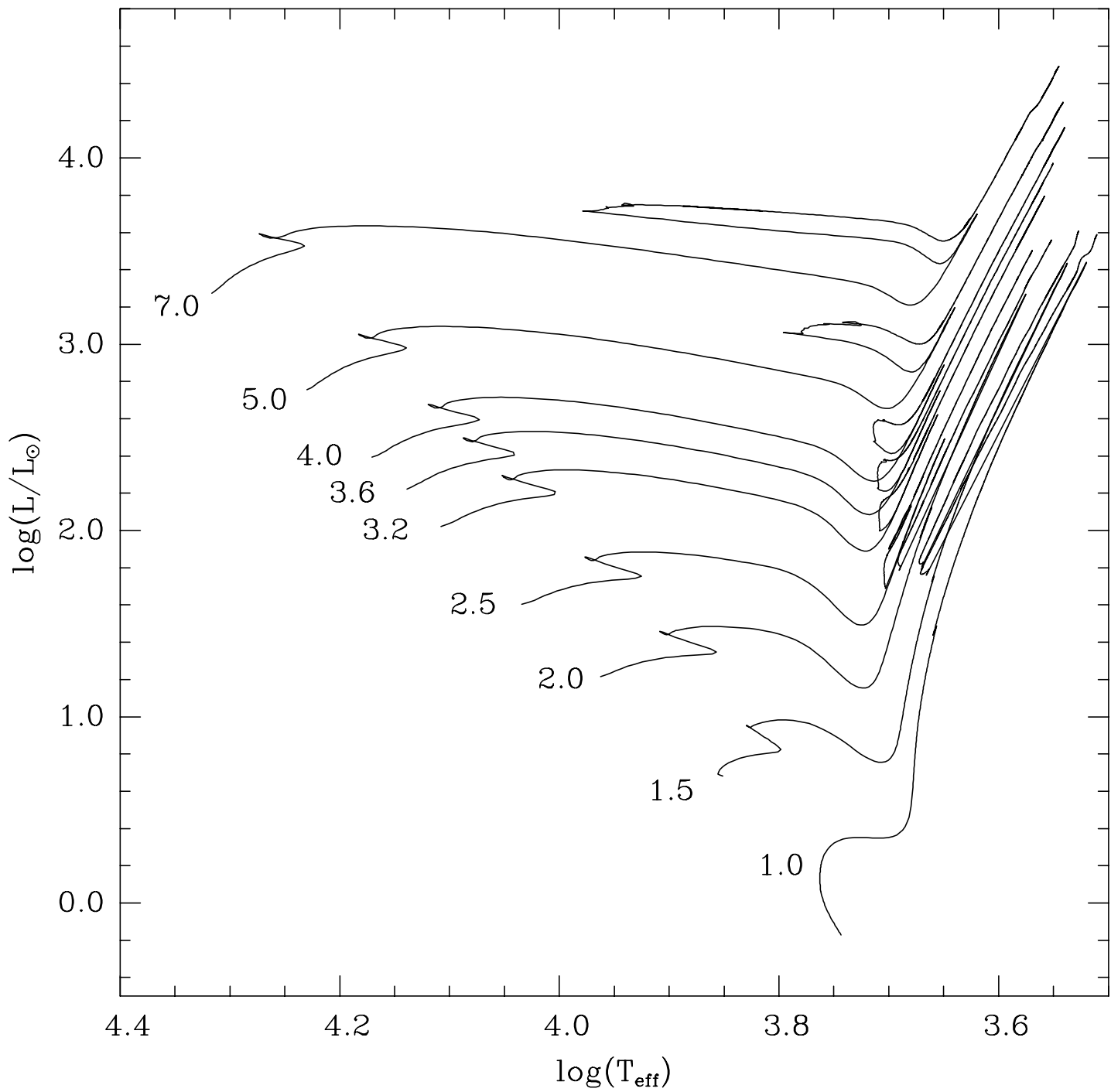
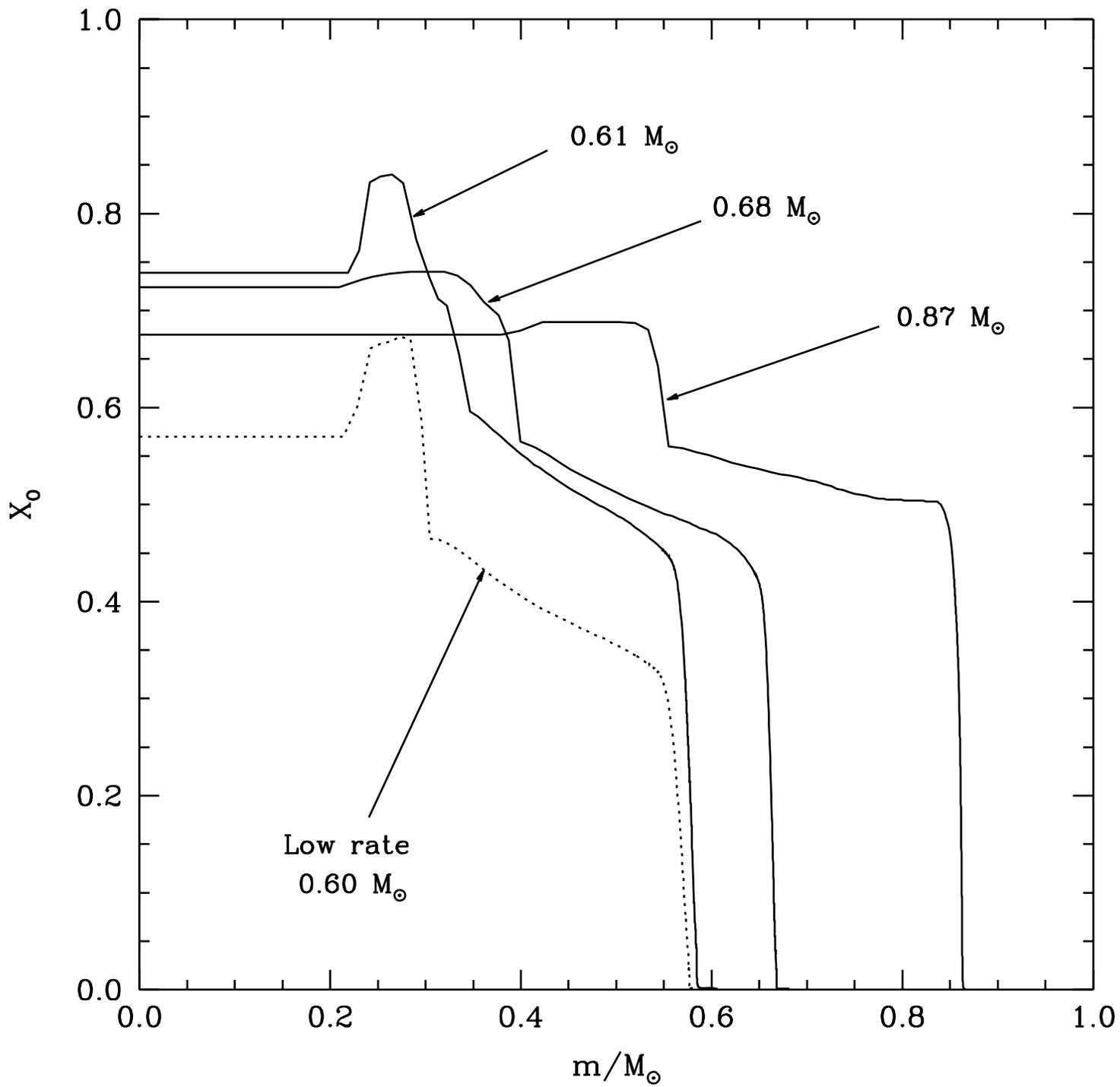
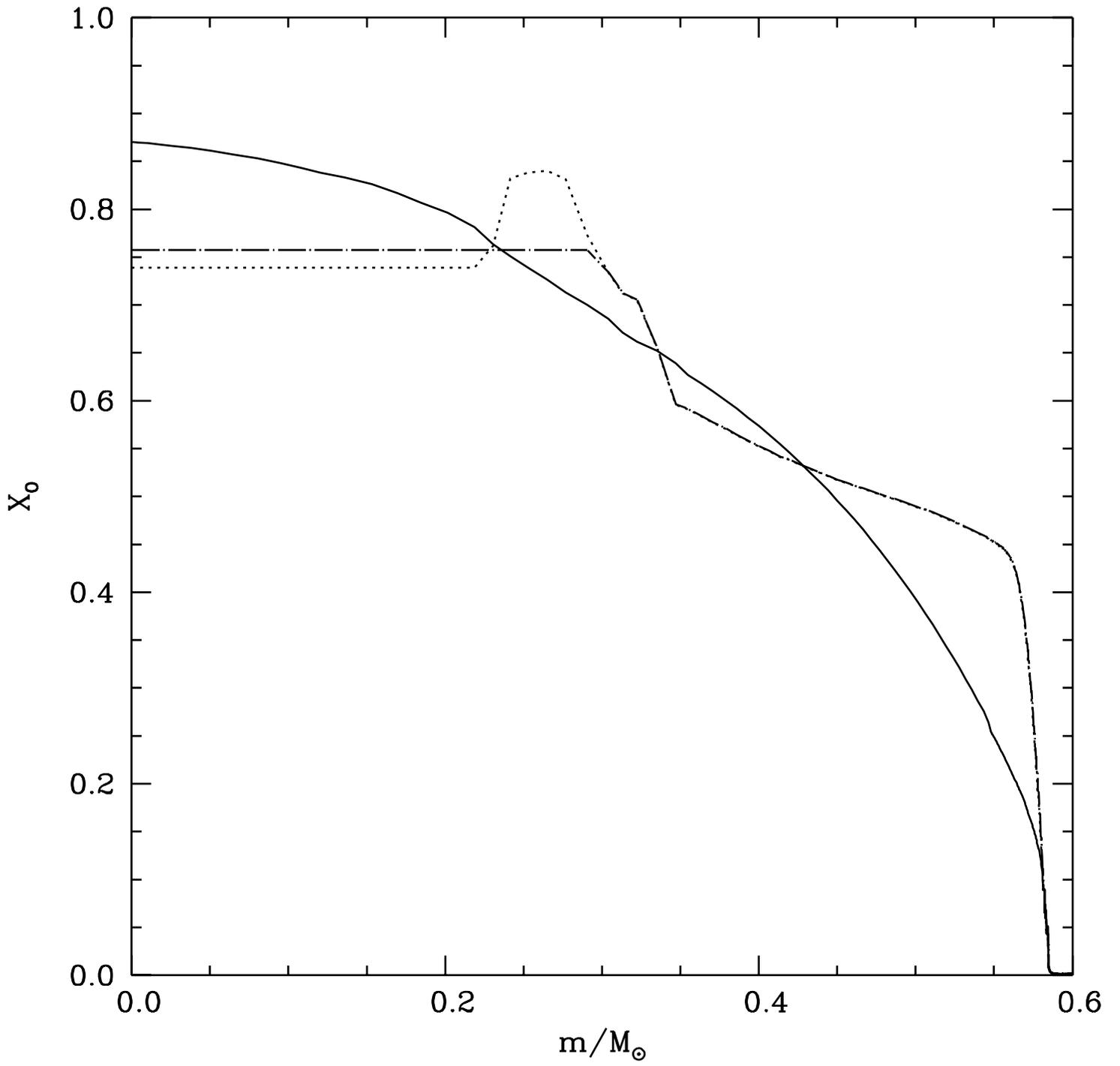
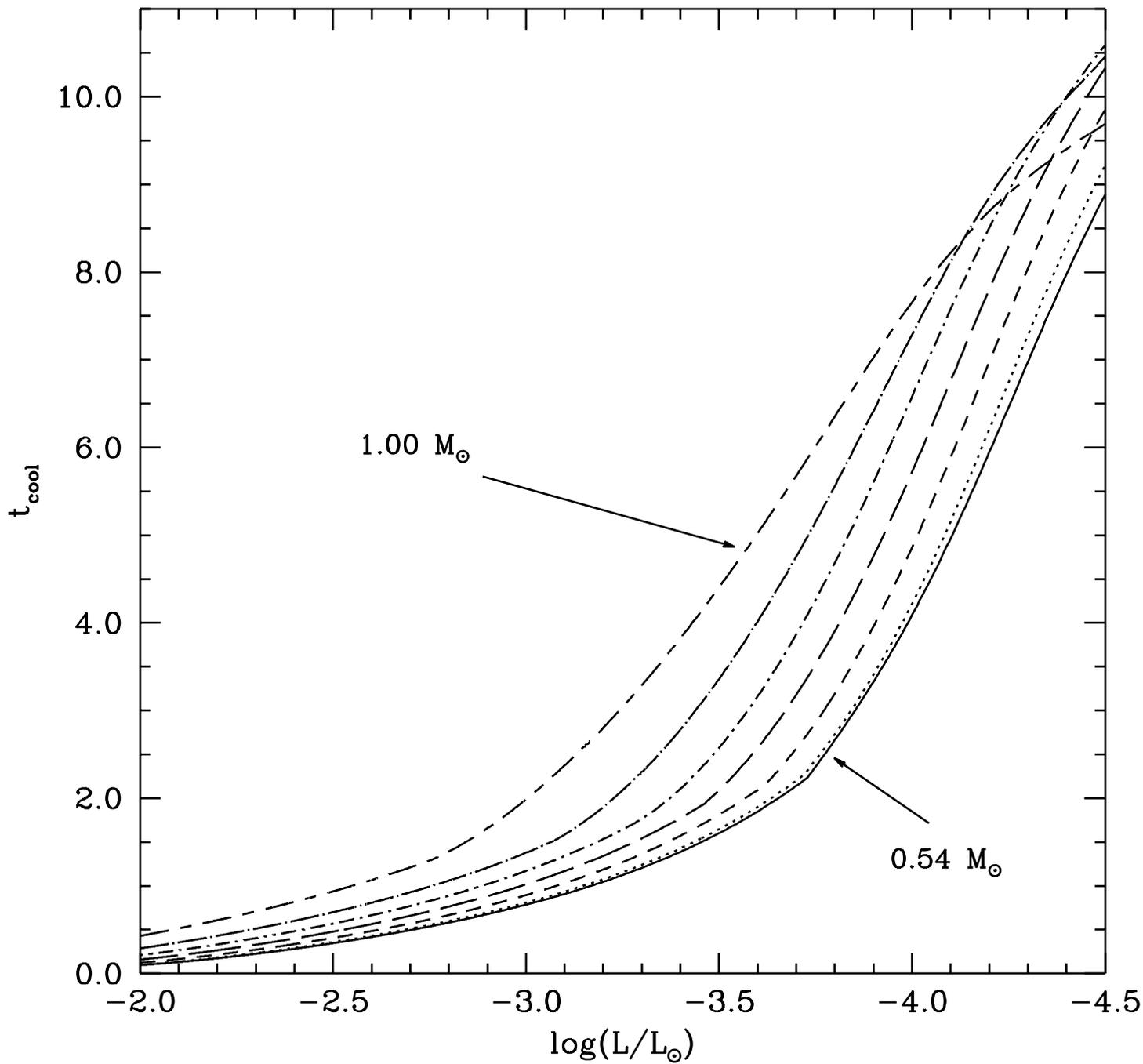
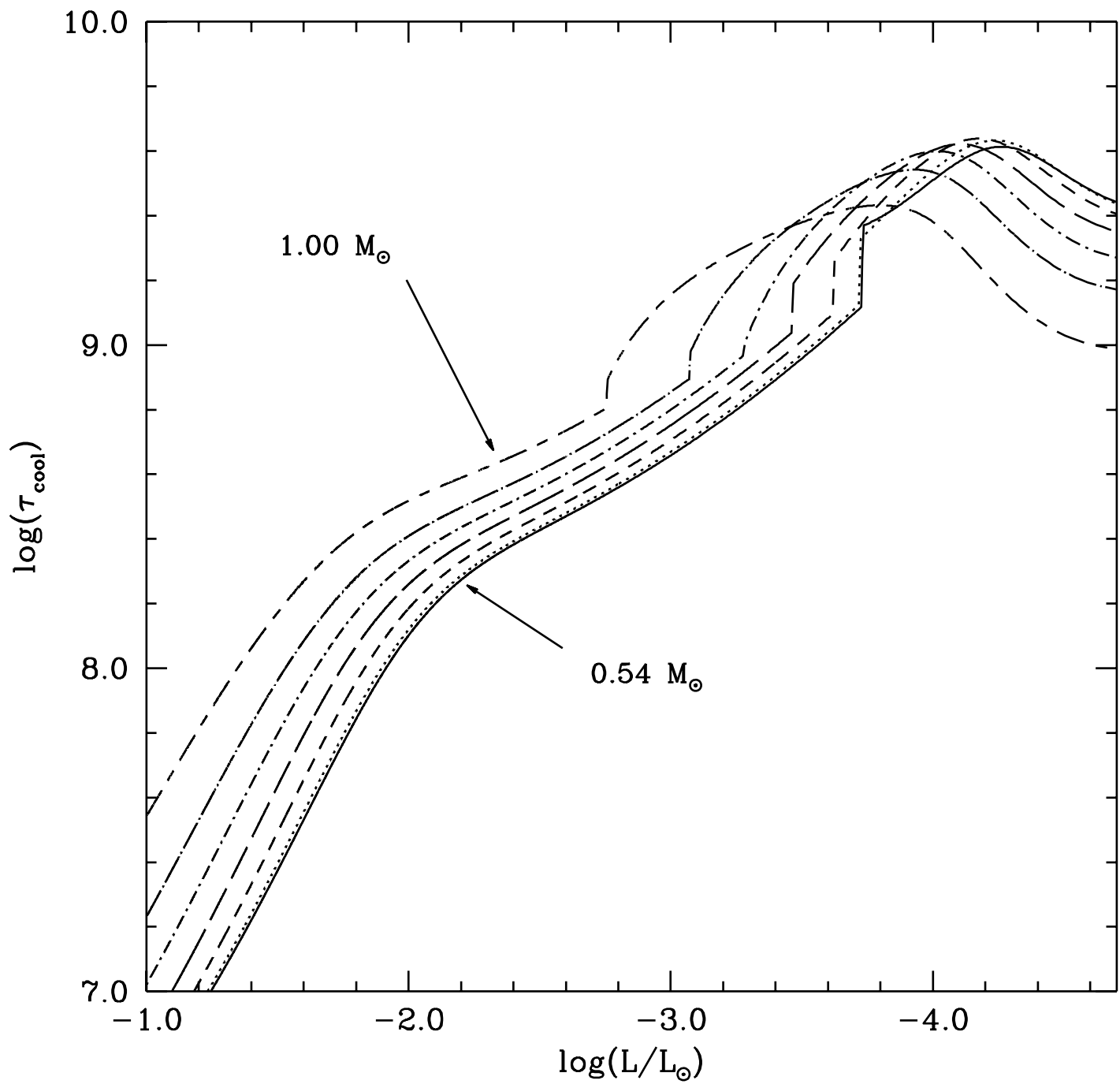
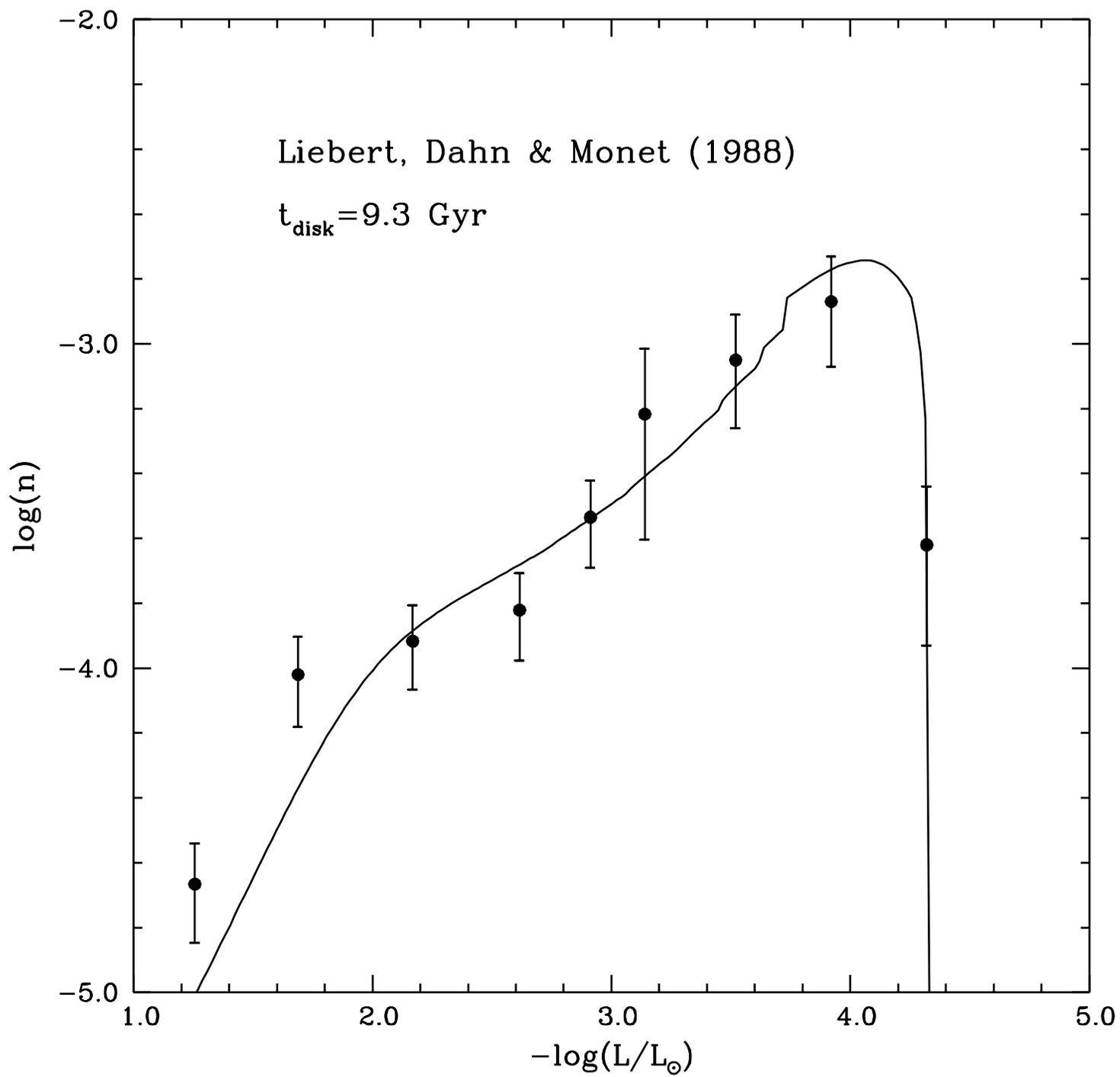
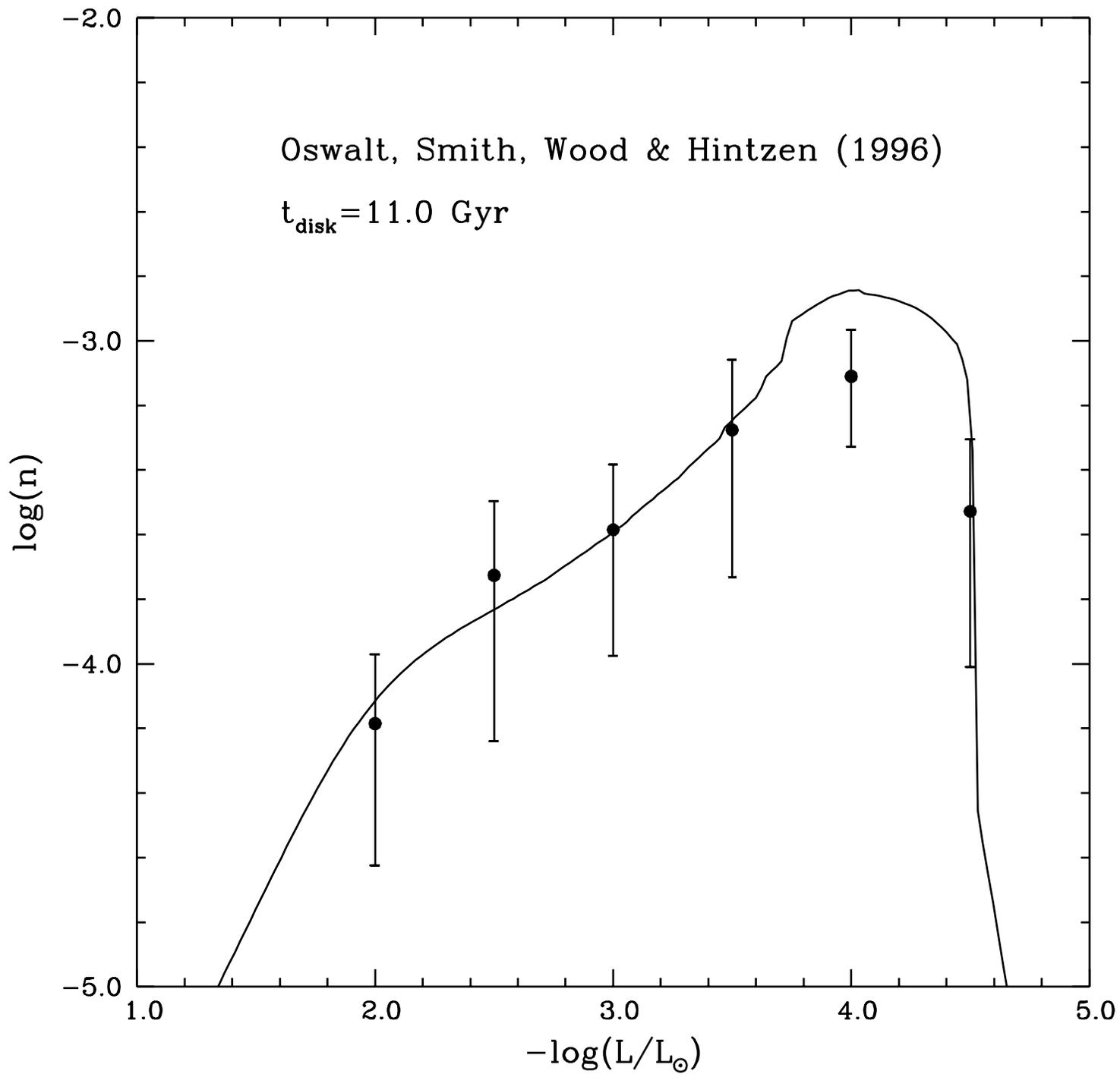
\end{document}